# Algebraic algorithms for multiloop calculations
# The first 15 years. What's next?

Fyodor V. Tkachov

*Institute for Nuclear Research of Russian Academy of Sciences, Moscow 117312*

The ideas behind the concept of algebraic ("integration-by-parts") algorithms for multiloop calculations are reviewed. For any topology and mass pattern, there exists a finite iterative algebraic procedure which transforms the corresponding Feynman-parameterized integrands into a form that is optimal for numerical integration, with all the poles in $D-4$ explicitly extracted.

**The Concept**

*A victory has a hundred fathers*

The concept of algebraic algorithms for multiloop integrals [1]–[4] has brought about an industry of analytical multiloop calculations[a], so it would not be inappropriate to reminisce upon the anguish of its birth [1], given a non-negligible amount of research funds being misallocated through misinformation[b].

Back in 1979, when perturbative QCD, IBM, and the USSR were in full bloom, the first project which I took part in was successfully completed. Our analytical calculation was receiving a considerable attention, and my contribution to it was such that there was no doubt in my mind that it would be my Ph.D. thesis. However, the senior member of our team of three intimated to me that the result would go instead into another thesis that would not be able to get composed "otherwise" (which was true beyond any shade of doubt). The deal (in which I had no negotiation powers) was accompanied by a swift promotion of the senior member.[c] A concrete reward for me was a deterioration of my eyesight by two diopters as a result of my innocent calculational enthusiasm. Moreover, due to the rigidity of Soviet system and my then insignificance, I was stuck in that vulnerable position for what seemed to be forever, although I was too light-hearted and inexperienced at the time to foresee all the consequences (which may have been just as well).

But research *was* fun and the next level of complexity (3-loop finite parts of massless self-energy diagrams) seemed within grasp. The appetites whetted, it was suggested that "we" tackle the many multiple series that emerged there within the technique that I had shown to be successful with 2-loop finite parts. However, I balked: if one were to appropriate the fruits of my meditations, one could at least leave me alone in the joy of the process. (That tacit arrangement was adhered to by the other side with much gusto all the way through the theory of asymptotic expansions, which is another story [7].)

Being thus vigorously motivated, I conceived of a method that would, ideally, meet the following criteria:

▶ My method would reduce human intervention to an absolute minimum. First, because my eyesight was at stake. Second, although I remember being able (then) to expand $\Gamma$-functions in 2-loops pretty reliably, I was less sure about 3 loops.

▶ That meant computer algebra and SCHOONSCHIP [8], so that my ideal method would be basically simple to make a program implementation feasible.

▶ Third, the usual approach was (and largely remains) to do all basic integrals "analytically" (i.e. by hand in the usual human fashion) and leave to computer only traces, scalar products, and substitutions of answers for the basic integrals. Since there were so many different integrals in 3 loops, I decided that such an approach was unacceptable: My method would have to do something about all those integrals too.

▶ Lastly, my ideal method — *my baby* — had to be *beautiful*. That was the only kind of consolation I could expect to obtain with any degree of certainty.

(**NB** The list is by no means obsolete after 15 years.)

Where did I have to look for such a method? Whenever one has to start from scratch (which one has to do surprisingly often [7], [9]), a little philosophizing at the level of first principles is appropriate (because nothing works like a sound first principle implemented with due simplicity and relentless logic):

One general consideration was not unknown (see e.g. [10]) but rarely heeded, namely, that best algorithms for

---

[a] A number of papers in the proceedings of all AIHENP workshops discuss calculations that make an essential use of identities obtained via differentiations in momentum space.

[b] often undocumented; but cf. an unsuspecting citation in [5].

Of course, Germany (say) is rich enough to afford some overhead, especially if one takes into account the grant-fetching value of the Russian multiloop expertise. However, "𝔚𝔥𝔬𝔰𝔬 𝔦𝔰 𝔭𝔞𝔯𝔱𝔫𝔢𝔯 𝔴𝔦𝔱𝔥 𝔞 𝔱𝔥𝔦𝔢𝔣 𝔥𝔞𝔱𝔢𝔱𝔥 𝔥𝔦𝔰 𝔬𝔴𝔫 𝔰𝔬𝔲𝔩: 𝔥𝔢 𝔥𝔢𝔞𝔯𝔢𝔱𝔥 𝔠𝔲𝔯𝔰𝔦𝔫𝔤, 𝔞𝔫𝔡 𝔟𝔢𝔴𝔯𝔞𝔶𝔢𝔱𝔥 𝔦𝔱 𝔫𝔬𝔱." [6]

[c] Human nature is, of course, much the same everywhere in the world. But one ought to remember about behavioral and social renormalization factors — in the present case, roughly, Mr. James Carter over *tovarishch* Leonid Ilyich [Brezhnev].

computer algebra applications should avoid imitating human ways of doing formulas. A rule of thumb is, one has to trade the true complexity of the amazing pattern matching that human beings are capable of, for a dull hugeness of more or less uniform data structures (e.g. polynomials) that computers are so good at. A difficulty here is that one has to understand quite well what's going on inside computer algebra systems. Another difficulty is in imagining an "analytical" algorithm that one may not even be able to perform oneself.

Furthermore, "calculation" is nothing but a "transformation". In order to "transform" an object one has to understand its "structure". What is "structure" of 3-loop massless self-energy integrals? (Exercise Answer this before reading on.) First of all, it is the set of identities expressing momentum conservation at vertices. That is already used in simplifications of scalar products in numerators. Next, "masslessness" means that Euler's identity $Q\partial f(Q) = \lambda f(Q)$ holds for massless self-energies with $Q$ the external momentum. (At that point I already knew better to be bothered by comments about how it was silly to attempt to do integrals by differentiations.) Doing differentiation on the integrand and simplifying scalar products one obtains an identity connecting several different integrals. The identity depends on how $Q$ is passed through the propagators, so I started writing out all possible variants and trying to combine them in all sorts of ways. It would have been easier to deal with 2-loop diagrams, but I did not know the answer and chose the 3-loop ladder as my playing ground, so it took several months to stumble upon the celebrated triangle relation [1]. That took care of planar diagrams. A little more effort yielded me non-planars in terms of one basic non-planar diagram (as well as a rebuke for having failed to explain sooner the entire significance of differential identities). I had a complete algorithm but could not resist the temptation of wasting some more time discovering (once a week or so) ways to reduce the one remaining scalar non-planar integral to planar ones before the algorithm was submitted for publication.

But that was not the whole story. When actual programming of the algorithm based on pure recursions [1], [2] was attempted, it proved a failure (the huge program simply would not work as a whole). The project remained at a standstill until an explicit solution of the triangle recursion was found [3]. It resulted in a great speedup in both debugging and performance of the program which finally began to turn diagrams into numbers in a way that caused me to baptize it "MINCER". I emphasize a central role of that solution in the actual functioning of MINCER [4] because misleadingly few references have been made to the fact in the literature.

———————————

**What's Next?**

> …For the duration of 40 hours he was expounding to Messrs. Gentlemen a great project, the execution of which subsequently brought a great glory to England and showed how far the human mind can reach out sometimes!
> The Drilling Through The Moon With A Colossal Auger
> — such was the subject of Mr. Lund's speech!… [11]

The range of applications of algebraic algorithms of the conventional type (differential identities in momentum space [1]) continues to expand (cf. the recent applications to the electron magnetic moment [12] and to the effective heavy quark theories [13]). What I'd like to discuss now is the case of integrals with several external parameters ($s$, $t$, masses etc.), which is vital in modern high energy physics applications.

First of all, with several parameters in the problem, the variety of analytical details is so enormous that one is driven by desperation and logic to ignore them all and to consider calculation of an *arbitrary* multiloop integral — however preposterous such an idea may seem!

Second, one has to be clear about the meaning of "calculation". The enormity of the problem forces one to interpret it in a strictly pragmatic fashion as obtaining a family of "satisfactorily fast" algorithms to produce required curves. Such algorithms need not be based on "analytical" formulas. In fact, with many parameters in the problem, numerics at the final stage can hardly ever be avoided.

Lastly, the true purpose of algebraic identities is not so much to "calculate" the integrals as to make them simpler OR reduce their number.[d]

**NB** The meaning of "simpler" is, again, strictly context-dependent: in the case of numerical integration it need not mean "fewer terms" in the integrand but e.g. "more continuous derivatives". For instance, the new multidimensional integration package MILX [15] provides an option of using higher-order quadratures along with Monte Carlo, thus greatly improving performance when integrands are sufficiently smooth.

So, let $x = (x_1,\ldots,x_K)$ be the vector of Feynman parameters; $S$ is the integration region, a simplex described by $x_j > 0; \sum_j x_j < 1; j = 1,\ldots,K$ ($K \equiv \dim S$); $Q(x)$ and $V_i(x)$ are arbitrary polynomials of $x$ with symbolic coefficients; $\mu_i$ are symbolic (complex, transcendental) parameters. Consider integrals of the form

---

[d] Contrast this with the "algebraic calculability" of [14] which attempts to normalize the spectacular success achieved in the case of massless self-energies which was merely a matter of luck. A situation much more like what I envisaged in the beginning occured in the case of anomalous magnetic moment where everything reduces to 18 independent integrals [12].

$$\int_S dx\, Q(x) \prod_i V_i^{\mu_i}(x). \quad (1)$$

Dimensionally/analytically regulated diagrams are obtained by appropriate choices of $Q(x)$, $V_i(x)$ and $\mu_i$. $\mu_i$ are equal to a fixed integer (denoted $n_i$; it can be positive, negative or zero) plus a symbolic part (denoted $\varepsilon_i$): $\varepsilon_i$ are regarded as infinitesimal because an expansion in $\varepsilon_i$ has to be performed ultimately. The point is that, whereas the integral (1) may be well-defined as a meromorphic function of $\mu_i$, the expansion in $\varepsilon_i$ may not be performed directly in the integrand because some $n_i$ may be too large and negative.

The structure I propose to exploit can be described as "polynomials to symbolic powers" and is fully specified by (1): no further information about $V_i(x)$ will be used. The key result I would like to present is as follows:

**Theorem A ("algorithm of algorithms of algebraic multiloop calculations").** *Any integral (1) can be reduced — via a finite number of purely algebraic steps — to the following form with any given integer $\omega \geq 0$:*

$$\sum_{S'} \frac{1}{B'} \int_{S'} dx'\, Q'(x') \prod_i V_i^{\omega+\varepsilon_i}(x'), \quad (2)$$

*where the sum contains a finite number of terms; $S'$ can be either $S$ or any of its boundary simplices with various $\dim S' \leq S$ ($\dim S' = 0$ is allowed); $x'$ parameterize $S'$ in a natural fashion (e.g. some sets $x'$ are obtained by nullifying certain components of $x$); $Q'(x')$ are polynomials of $x'$; $B'$ are independent of $x'$; $B'$ and all coefficients of $Q'(x')$ are polynomials of all $\mu_i$ and of the coefficients of $Q(x)$ and $V_i(x)$; $V_i(x')$ are restrictions of $V_i(x)$ to the corresponding boundary simplices $S'$; all numeric coefficients involved are integer.*

The proof uses the following result:

**Theorem B ("generalized Bernstein functional equation").** *For any finite set of polynomials $V_i(x)$ there exists an identity of the following form:*

$$B^{-1}\mathcal{P}(x,\partial) \prod_i V_i^{\mu_i+1}(x) = \prod_i V_i^{\mu_i}(x), \quad (3)$$

*where $\mathcal{P}(x,\partial)$ is a polynomial of $x$ and $\partial_i = \partial/\partial x_i$; $B$ and all coefficients of $\mathcal{P}$ are polynomials of $\mu_i$ and of the coefficients of $V_i(x)$.*

Proof of Theorem B   Using methods of abstract algebra ("finitely generated ideals" of "graded left modules", etc.), one can establish (see e.g. [16]) existence of identities (3) for the case of exactly one (arbitrary) polynomial — a result known as the Bernstein functional equation:

$$b^{-1}\overline{\mathcal{P}}(x,\partial) V^{\mu+1}(x) = V^\mu(x), \quad (4)$$

where both $b$ and $\overline{\mathcal{P}}$ depend polynomially on $\mu$ and on the coefficients of $V$. Take $V(x) = \prod_i V_i(x)$, perform the differentiations in (4), cancel the symbolic powers and equate coefficients of monomials of $x$ on both sides; one obtains a linear system of equations for the unknown coefficients of $\overline{\mathcal{P}}$. Bernstein's result is equivalent to existence of a non-zero solution of the system, i.e. some determinant is non-zero. Perform a similar reduction for (3), and compare the resulting linear system with that obtained for (4). The former differs from the latter by replacements of different entries of $\mu$ by different $\mu_i$, so that the determinant mentioned above cannot be zero. Therefore, the system has a non-zero solution, which yields the identity (3) with the coefficients of $\mathcal{P}$ that are rational functions of both $\mu_i$ and the coefficients of $V_i(x)$. Taking out all the denominators and combining them into $B$ yields (3).   □

Proof of Theorem A   One iteratively performs replacements of the integrand with the corresponding l.h.s. of (3) and performs integrations by parts to get rid of the differentiations. One does this (with appropriate $\mathcal{P}$ and $B$) also for all integrals corresponding to the boundary terms resulting from integration by parts. After a finite number of iterations the integer parts of all complex powers become $\geq \omega$.   □

In dimensionally regulated diagrams, all $\mu_i$ are linear combinations of one symbolic parameter $\varepsilon$. This may cause problems due to nullification of some $B$. That would mean a breakdown of dimensional regularization so that an additional analytic regularization would have to be used to make all $\mu_i$ independent. But since the identities (3) have to be found for arbitrary $\mu_i$ anyway (to handle integrals with different integer parts of $\mu_i$), there is no practical problem here.   ■

**Discussion**

- The operators $\mathcal{P}$ for the same topology but different mass patterns need not be connected in any simple way.
- There is an infinite family of non-trivially different operators $\mathcal{P}$ satisfying (3) for the same $V_i(x)$.
- The only way, known to me at the time of this writing, of finding the polynomials $\mathcal{P}$ is via a direct study of the linear system of equations for the unknown coefficients of $\mathcal{P}$ (cf. the proof of Theorem B). This is cumbersome (the form of the system depends on the degree of $\mathcal{P}$ w.r.t. derivatives), but one has to find such an identity only once for each topology/mass pattern, and only polynomial algebra with rational coefficients is involved.
- Multiplying both sides of (3) from the left by some of $V_i(x)$'s and absorbing the latter into $\mathcal{P}$ on the l.h.s., one

proves existence of identities with only a subset of exponents raised by +1, with others intact. In general, the family of operators of this latter form contains ones that cannot be obtained from (3) as just described. These additional identities may result in more economical algorithms (iteratively raising each symbolic power to exactly the minimal value required for achieving the desired smoothness of the integrand).

- Similarly, one can directly seek an identity with +1's on the l.h.s. replaced with $+N_i$'s with required $N_i \geq 0$ (or even with some $N_i < 0$ if the integer part of the $i$-th original power was already positive), and with $Q$ on the r.h.s. This is similar to explicit solution of recurrent relations in the conventional algorithms [3], [12].

- More iterations make integrands smoother, but the original integrals did have singularities in external parameters! The paradox is resolved by noticing that the denominators $B$ have zeros. Those zeros, therefore, *must* correspond to both threshold singularities (hence $B$ must be connected to determinants of the system of Landau equations) *and* to the poles in $D-4$ (in the case of dimensional regularization).

- It follows that the algorithm automatically extracts the poles in $D-4$: just expand integrands in the end in $D-4$; complex powers give rise to integer $\geq 0$ powers of $V_i(x)$ and $\log V_i(x)$; all the poles come from $B$'s. With well-defined integrals for the coefficients of the poles, their cancellations may be checked numerically.

- Due to zeros of $B$'s, the numerical advantage of smoother integrands is lost near thresholds (where numerical convergence is poor anyway) due to cancellations in the sum (2). However, the regions near thresholds may be systematically treated via asymptotic expansions obtained by the method of asymptotic operation [7].

- The identities of the original "integration-by-parts" algorithm [1]–[2] can be regarded as special cases of (3). In that sense, Eq. (3) represents the most general basic type of algebraic identities for multiloop integrals, which justifies the name of Theorem A. Concerning the identities connecting integrals with different values of the space-time dimension $D$ (cf. [18], [20], [19]), one notices that $D$ enters exponents of different factors with *different* signs, whereas in (3) all exponents change in the *same* direction. So one cannot naturally interpret the new identities as recurrences w.r.t. $D$ (beyond one loop).

- As a simple application, consider dimensionally regulated one-loop integrals which contain just one complex exponent, $\int_S \mathrm{d}x \, Q(x) V^{-n-\varepsilon}(x)$, where $n$ is a non-negative integer, $\varepsilon$ is the complex regularization parameter, and $V(x)$ is a *quadratic* polynomial of $x$. Without loss of generality $V(x) = x^{\mathrm{T}} \widetilde{V} x + 2R^{\mathrm{T}} x + Z$ where $\widetilde{V}$ is a $K \times K$ matrix and $R$ is a $K$-vector. Then

$$\frac{1}{\Delta}\left(1 - \frac{(x+A)\partial_x}{2(\mu+1)}\right) V^{\mu+1}(x) = V^{\mu}(x), \qquad (5)$$

where $\Delta = \left(Z - R^{\mathrm{T}}\widetilde{V}^{-1}R\right)$ and $A = R^{\mathrm{T}}\widetilde{V}^{-1}$. (Special cases of this relation in integral form have been known for some time; cf. [18].) (**Exercise** Write this out for your favorite one-loop integral.) (**NB** If $\widetilde{V}$ is singular then one should determine its null space and treat transverse and longitudinal coordinates separately.) Usually, one first reduces non-scalar one-loop integrals to scalar ones and then uses formulas for the latter in terms of dilogarithms etc. — a typical example of a procedure motivated by how human beings perform calculations. The computer-oriented approach based on (5) is to apply it iteratively to the integrand (with all the numerators) until required regularity is achieved; then one can take advantage of the numerical integration schemes that employ higher-order quadratures such as MILX [15].[e]

- For 2-loop diagrams, the situation is as follows. For the simplest sunset topology, if $\mathcal{P}$ is second order w.r.t. $\partial$'s (a minimal possible number of derivatives that yields a non-trivial identity (3)) then $\mathcal{P}$ has 5-th powers of $x$'s. (**NB** The maximal powers of $\partial$'s and $x$'s are anticorrelated.) This is quite cumbersome, but it is a dull complexity of a polynomial.

- Being a monotonous repetition of the same cumbersome algebraic identity, the algorithm is as unfitted for hand calculations as it is perfect for programming.

**Conclusions**

It took about 10 years for MINCER [4], an implementation of the first algebraic multiloop algorithm [1]–[3], to deliver reliable results. The amount of algebra involved in the construction and iterative application of (3) is even greater, and may prove huge enough to prevent the new "algorithm of algorithms" from being actually used beyond one loop to any greater extent than the SONG OF SONGS is sung.

But consider this: If an identity (3) is explicitly constructed for a given topology/mass pattern, then it *simultaneously* does the following things — each of which would normally be an achievement per se:

▶ *It reduces integrands to a form that is optimal for numerical integration.*

▶ *It allows straightforward program implementation.*

▶ *It explicitly extracts* all (!) *the poles in $D-4$.*

Taking into account the unrestricted generality of the new approach (it is applicable — at least in theory — to *any* multiloop integral, including ones with, say, non-relativistic propagators etc. etc.) and the pace of progress of computer industry, it looks like it *might* possess a potential to relieve us from some grueling chores in future and to give a boost to the industry of loop calculations (which needs it [21]; remember also the funds that went into the lattice QCD). The operators $\mathcal{P}$ could be found by solving a system of linear algebraic equations for their coefficients — a highly cumbersome procedure given all the symbolic parameters involved ($D$, masses etc.) but requiring nothing that would go much beyond the standard arsenal of polynomial computer algebra. So, anyone equipped with a sufficiently fast and reasonably flexible computer algebra system may already try one's hand at constructing an identity (3) for one's favorite two-loop topology.

---

[e] A specialized version of MILX is to be made available.

Acknowledgments The trigger for this work was a conversation with K. Kato at the Xth Zvenigorod workshop (INP MSU, September, 1995) concerning numerical computation of loop integrals within the framework of the GRACE project [22]. I thank N. Sveshnikov for a discussion, and J. Vermaseren, D. Perret-Gallix and V. Ilyin for helping a signal to get through. Financial support came from AIHENP'96, ISF-Logovaz and the Russian Foundation for Basic Research (grant 95-02-05794).